\documentclass[pre, onecolumn,showpacs,superscriptaddress,aps]{revtex4}
\usepackage{amsfonts}
\usepackage{amsmath}
\usepackage{amssymb}
\usepackage{graphicx,float}
\usepackage{dcolumn}
\usepackage{times}
\usepackage{color}

\newcommand{\beq}[0]{\begin{equation}}
\newcommand{\eeq}[0]{\end{equation}}
\newcommand{\e}{\varepsilon}
\newcommand{\thet}{\vartheta}

\newcommand{\w}{\omega}
\newcommand{\bc}{\begin{center}}
\newcommand{\ec}{\end{center}}

\begin{document}

\title{NLS approximation and dark solitons for the Adlam-Allen model
of cold collisionless plasmas}
\author{G.P. Veldes}
\affiliation{Department of Physics, University of Thessaly, Lamia 35100, Greece}
\affiliation{Department of Physics, University of Athens, Panepistimiopolis,
Zografos, Athens 15784, Greece}
 \author{V. Koukouloyannis}
\affiliation{Department of Mathematics, University of the Aegean, Karlovassi, 83200 Samos, Greece}
\affiliation{Department of Mathematics and Statistics, University of Massachusetts, Amherst MA 1003-4515, USA}
\author{D.J.\ Frantzeskakis}
\affiliation{Department of Physics, University of Athens, Panepistimiopolis,
Zografos, Athens 15784, Greece}
\author{P.G.\ Kevrekidis}
\affiliation{Department of Mathematics and Statistics, University of Massachusetts, Amherst MA 1003-4515, USA}

\begin{abstract}
The present work extends earlier considerations 
on a quintessential model of cold, 
collisionless plasmas, namely the Adlam-Allen model. 
Previously, an analysis of homoclinic solutions around
a non-vanishing background (associated with a saddle 
equilibrium) led to a Korteweg-de Vries reduction. Here, we consider
a different equilibrium of the co-traveling frame model,
namely a center, and expanding around it, by means
of a multiple-scale methodology and suitable scalings, leads to an effective 
defocusing nonlinear Schr{\"o}dinger 
equation.
Leveraging the latter, we construct, for the first
time to our knowledge, physically realistic
dark soliton waveforms of the Adlam-Allen model.
We subsequently test the numerical evolution of such
coherent structures, identifying them as long-lived waveforms of the full model. 
\end{abstract}


\maketitle
\section{Introduction }

In the early stages of the development of nonlinear
science, a number of models captivated the interest
of the emerging nonlinear wave community, especially
through the explosion of interest in integrable systems.
This occurred through models such as the Korteweg-de Vries (KdV) equation~\cite{zabusky1965interaction} (which
reemerged in nonlinear lattices~\cite{fermi1955studies}, 
long after it was considered in water waves~\cite{korteweg1895change}), as well
as the nonlinear-Schr{\"o}dinger (NLS) equation
which was proposed in the context of
nonlinear optics~\cite{townes}, but also 
used early on 
in the context of water waves studies~\cite{zakharov}.

One of the contexts where such classes of dispersive
nonlinear PDE models emerged in these first steps
was in the study of plasmas, e.g., through the classic
work of~\cite{washimi1966propagation}, exploring
the propagation of ion-acoustic traveling waves of
small amplitude, via the KdV equation.
However, what seems to be far less well-known
was that almost a decade earlier than this classic
work, another remarkable model encompassing hydrodynamic
waves in plasmas was proposed by Adlam and Allen (AA)
in 1958~\cite{Allen1958} and was subsequently expanded
upon in further work in 1960~\cite{adlam1960}, 
preceding by 7 years the birth of soliton theory
through the classic study of~\cite{zabusky1965interaction}.
The physical setting of the
AA model featured electrons and ions in a plasma
with a magnetic field component along the $z$-direction
(but only dependent along the spatial variable $x$).
Assuming that considerable energy was imparted
to the particles in the waves~\cite{Allen1958}, this
model was intended towards potential applications
in fusion research, as well as in astrophysical 
phenomena, related to wave phenomena in space and solar system physics 
\cite{meyer-vernet2007basics}. 

Over the past few years, there has been a revival of
interest in the AA model. On the one hand, this has been
due to the interest in obtaining a full description
of the particle properties and electro-magnetic
fields, as was done in~\cite{Abbas2020} (and for
a different configuration of electric and magnetic
fields in~\cite{Abbas2022}). On the other hand, this
has stemmed from an effort to analyze the mathematical
properties of the model and, especially, of its
solitary and periodic waves, e.g., through a multiscale
expansion leading to the KdV equation and connecting
the AA waves with their KdV counterparts~\cite{Allen}, as well
as through a study of their pairwise interactions~\cite{koukouloyannis2022interactions}.

In the present work, we revisit the  prototypical
form of the AA model and illustrate a completely
unexplored family of waveforms that the model
can manifest. We are motivated by the analysis of the
co-traveling frame where the well-known solitary
waves of the model are found to exist as homoclinic
orbits to a saddle-point in the resulting phase plane.
However, as was already seen in~\cite{Allen}, in addition
to the saddle point, the relevant phase portrait also
features a center. Around this center, periodic orbits
are natural to exist and modulations within the envelope
of these periodic orbits can be expected to formulate
an entirely unprecedented {\it dark soliton}
structure of the AA equations. To showcase this
phenomenology, we develop a multiple-scales expansion
of the small amplitude solutions around the center,
a feature that, given the nonlinearity of the model,
enables us to obtain at the suitable order an NLS
model; indeed, and while connections of AA to KdV
have been discussed in the past, we are not aware
of any such prior connections to NLS. Then, realizing
the defocusing character of the resulting NLS and 
``borrowing'' from there the general dark solitonic (DS)
waveform,  we utilize the reconstruction based on the
multiscale expansion to formulate and numerically 
test this modulated dark solitary wave within the
original AA model. We find good agreement with the
expectation from our analytical results and, indeed,
thus infer the long-lived nature of the DS pattern in the
cold, collision-free plasma. The limitations of
the approach are also analyzed. In an attempt to partially overcome these limitations, we also construct a hybrid dark-soliton solution of the system, considering a cnoidal background and applying on top of it an appropriate modulation based on the previously calculated solution. Then, we perform a numerical study of this solution and compare the results with the previous acquired ones.

Our presentation is structured as follows. In Section
II, we present 
the AA model and its setup.
In Section III, we delve into the multiple
scale analysis and offer an epigrammatic 
presentation of the
tedious calculations that lead to the NLS result. 
Finally, we turn to numerical computations where we
test our findings in Section IV. Here our numerical
results are shown for different values of the
intrinsic model, as well as perturbation parameters. Finally, we perform a numerical study of the system for the hybrid solution with the cnoidal background for larger amplitudes.
Finally, in Section V, we summarize our findings
and present our Conclusions.

\section{Adlam-Allen model and traveling wave solutions}

\subsection{Derivation of the AA model}
In its quintessential form for the description of
a cold, collision-free plasma permeated by a uniform magnetic field $\mathbf{B_0}=B_0 \hat{z}$, the AA model
examines the case where  a wave propagates in  the direction $x$, $\mathbf{k}= \hat{x}k$, perpendicular in the magnetic field $B_0$, into the undisturbed plasma with density $n_0$. 
First, we assume that the ion and electron densities are  $n_i\approx n_e$,  where the difference between $n_i$ and $n_e$ can be ignored except for their role in Poisson's equation (this is the so-called
quasineutrality of the plasma), with $j=i, e$ referring to ions and electrons, respectively. This small difference between the ion and electron densities produces an electrostatic field $\mathbf{E_x}=E_x \hat{x}$ in the direction of propagation. Due to the presence of an external magnetic field $B_0$ in the $z$ direction, transverse electron currents, as well as an electric field $\mathbf{E_y}=E_y \hat{y}$ in the $y$ direction, are simultaneously created. The increase in the magnetic field, as a disturbance of the magnetic field, $\mathbf{B_z}=B_z \hat{z}$, is due to the transverse currents.
Consequently, the velocity of electrons and ions have components only on the plane $(x,y)$, as does the electric field $\mathbf{E}=E_x \hat{x}+E_y \hat{y}$, while the magnetic field features solely 
a $z$-component ($\mathbf{B}=B_z \hat{z}$). 

%
To derive the AA model we start from Maxwell's equations: 
\begin{eqnarray}
\nabla \cdot \mathbf{E}&=&4\pi\rho,
\label{eq:max1}\\
\nabla \cdot \mathbf{B}&=&0,
\label{eq:max2}\\
\nabla \times \mathbf{E}&=&-\frac{1}{c}\frac{\partial \mathbf{B}}{\partial t},
\label{eq:max3}\\
\nabla \times \mathbf{B}&=&\frac{4\pi}{c} \mathbf{J}. 
\label{eq:max4}
\end{eqnarray} 
Here, $c$ is the velocity of light in vacuum, $\rho=n_iq_i +n_eq_e$ 
(with $q_i=-q_e=|e|$ being the electron and ion charges) is the charge density, and   $\mathbf{J}=q_in_i \mathbf{v}^{(i)}+q_e n_e \mathbf{v}^{(e)} $ (where $\mathbf{v}^{(i)}$ and $\mathbf{v}^{(e)}$ are the ion and electron velocities, respectively) is the 
current density. Maxwell's equations are coupled with the equations of continuity and motion for the electrons and ions:
\begin{eqnarray}
&&\frac{\partial n_{j}}{\partial t}+\nabla \cdot (n_j \mathbf{v}^{(j)})=0,
\label{eq:cont} 
\\
&&\frac{\partial \mathbf{v}^{(j)}}{\partial t}+(\mathbf{v}^{(j)}\cdot \nabla) \mathbf{v}^{(j)}  = \frac{q_j}{m_j}\left(\mathbf{E}+\mathbf{v}^{(j)} \times \frac{\mathbf{B}}{c}\right),
\label{eq:mot}
\end{eqnarray}
where $m_j$ is the mass (of ions or electrons). 
Additional assumptions involve the equality
of the $x$-components of the velocities of the two species, namely $v_x^{(i)}=v_x^{(e)} \equiv U_x$, while the $y$-components are considered to obey $v_y^{(i)}-v_y^{(e)} \equiv U_y$. Furthermore,  the  approximately equal 
ion and electron densities will be hereafter denoted by $n$.
Under the above considerations, Eqs.~(\ref{eq:max1})-(\ref{eq:mot}) are reduced to the form:
\begin{eqnarray}
&&\frac{\partial U_x}{\partial t}+U_x\frac{\partial U_x}{\partial x}=-\frac{|e| U_y B_z}{(m_i+m_e)c}
\label{eq:d1},
\\
&&\frac{\partial U_y}{\partial t}+U_x\frac{\partial U_y}{\partial x}=-\frac{|e|(m_i+m_e)}{m_i m_e}\left(E_y-U_x\frac{B_z}{c}\right),
\label{eq:d2}
\\
&&\frac{\partial n}{\partial t}+\frac{\partial (nU_x)}{\partial x}=0,
\label{eq:d3}\\
&&\frac{\partial B_z}{\partial x}=\frac{4\pi }{c}|e| n U_y,
\label{eq:m1}
\\
&&\frac{\partial E_y}{\partial x}=-\frac{1}{c}\frac{\partial B_z}{\partial t},
\label{eq:m2}
\end{eqnarray} 
and it is noted that it is the relative velocity $U_y$ (along the $y$-direction)  between electrons and ions that drives the relative motion.

Next, we express the above system in Langrangian coordinates $(h,t)$ (with $h$ being the spatial coordinate at $t=0$), 
\begin{eqnarray}
\frac{\partial U_x}{\partial t}&=&-\frac{|e| U_y B_z}{(m_i+m_e)c}
\label{eq:un1}\\
\frac{\partial U_y}{\partial t}&=&-\frac{|e| (m_i+m_e)}{m_i m_e}\left(E_y-U_x\frac{B_z}{c}\right)
\label{eq:un2}\\
\frac{\partial n}{\partial t}
&=&\frac{\partial U_x}{\partial h}
\label{eq:un3}\\
\frac{\partial B_z}{\partial h}&=&\frac{4\pi n_{0} |e| U_y}{c}
\label{eq:un4}\\
\frac{\partial E_y}{\partial h}&=&\frac{1}{c}\left(U_x\frac{\partial B_z}{\partial h}-\frac{1}{n}\frac{\partial B_z}{\partial t}\right), 
\label{eq:un5}
\end{eqnarray} 
%
%
%
and adimensionalize
the equations as follows. We measure the external magnetic field $B_z$ in units of $B_0$, 
the density $n$ in units of the unperturbed density $n_0$ 
[see Refs.~\cite{Allen1958,Allen})], 
the velocities $U_x, U_y$ in units of the characteristic Alfv{\'e}n velocity $V_A=\sqrt{ B_0^2/[4\pi n_0(m_i+m_e)]}$, 
the electric field $E_y$ in units of $E_0=V_A B_0/c$  
time $t$ in units of $t_0=(m_i m_e)^{1/2}c/(|e|B_0)$, 
 and the space variable $h$ in units $d=\sqrt{ c^2m_i m_e/[4\pi  n_0 e^2(m_i+m_e)]}$. 
This way, Eqs.~(\ref{eq:un1})-(\ref{eq:un5}) can be reduced to the following dimensionless partial differential equations (PDEs) for  the normalized magnetic field $B$ and inverse
density $R= 1/n$ (which is thus expected to physically satisfy $R>1$), in line with~\cite{adlam1960}:
\begin{eqnarray}
&&\frac{\partial^2 R}{\partial t^2} =-\frac{1}{2}\frac{\partial^2 B^2}{\partial x^2},
\label{eq:aa1} \\
&&\frac{\partial^2 B}{\partial x^2}=RB-1,
\label{eq:aa2} 
\end{eqnarray}
where the spatial coordinate $x$ corresponds to normalized $h$. As was shown in earlier works~\cite{Allen1958,adlam1960,Allen,Abbas2020}, the relevant traveling wave of interest to this
model occurs on top of the finite background $R=1, B=1$.
It is thus convenient to employ the linear transformation 
\beq R(x,t)=1+u(x,t), \quad 
B(x,t)=1+w(x,t), \label{eq:RB2uw}
\eeq 
and derive the following system of coupled nonlinear PDEs for the fields $u(x,t)$ and $w(x,t)$: 
%
\begin{eqnarray}
&&\frac{\partial^2 u}{\partial t^2}+ \frac{\partial^2 w}{\partial x^2}+\frac{1}{2}\frac{\partial^2 w^2}{\partial x^2}=0,
\label{eq:eq1} \\
&&\frac{\partial^2 w}{\partial x^2}- w- u-uw=0.
\label{eq:eq2} 
\end{eqnarray}
%

%

%

\subsection{Periodic  waves of the AA model}
\label{cnoidals}

Following the approach of~\cite{Allen}, we consider the co-traveling frame of reference, which is defined by the transformation $\xi=x-vt$, where $v$ is the constant velocity of the frame. We also assume that the two fields of the system depend only on $\xi$, i.e.,~ $u=u(\xi)$ and $w=w(\xi)$. This way, Eqs.~(\ref{eq:eq1})-(\ref{eq:eq2}) decouple, and one can derive an ordinary differential equation (ODE) for the field $w(\xi)$. Indeed, from \eqref{eq:eq1} we obtain
\beq
u=-\frac{1}{v^2}\left(w+\frac{w^2}{2}\right),
\label{eq:dsu}
\eeq
and substituting into \eqref{eq:eq2},
we obtain the following second-order ODE for $w(\xi)$: 
\beq
w_{\xi\xi}=\frac{v^2-1}{v^2}w-\frac{3}{2v^2}w^2-\frac{1}{2v^2}w^3.
\label{eq:dsw}
\eeq
This ODE can be viewed as the equation of motion, of the form $w_{\xi\xi}=-V'(w)$, of a unit mass particle in the presence of the potential $V(w)$,  given by: 
\beq
V(w)=-\frac{v^2-1}{2v^2}w^2+\frac{1}{2v^2}w^3+\frac{1}{8v^2}w^4.
\label{eq:V}
\eeq
Note that the requirement $v>1$ for the existence of a homoclinic orbit (corresponding to a solitary wave solution), as well as the requirement $R>1$ (ensuring that the density perturbation is always smaller than the equilibrium density), are fulfilled as long as the velocity $v$ is such that $1<v<2$.  
In this case, 
the potential possesses a local minimum, lying at 
$w_{min}=\frac{1}{2}(-3+\sqrt{8v^2+1})>0$,
for positive values of $w$ (which are the physically relevant values of $w$). 

\begin{figure}[tbp]
\begin{center}
\includegraphics[scale=0.3]{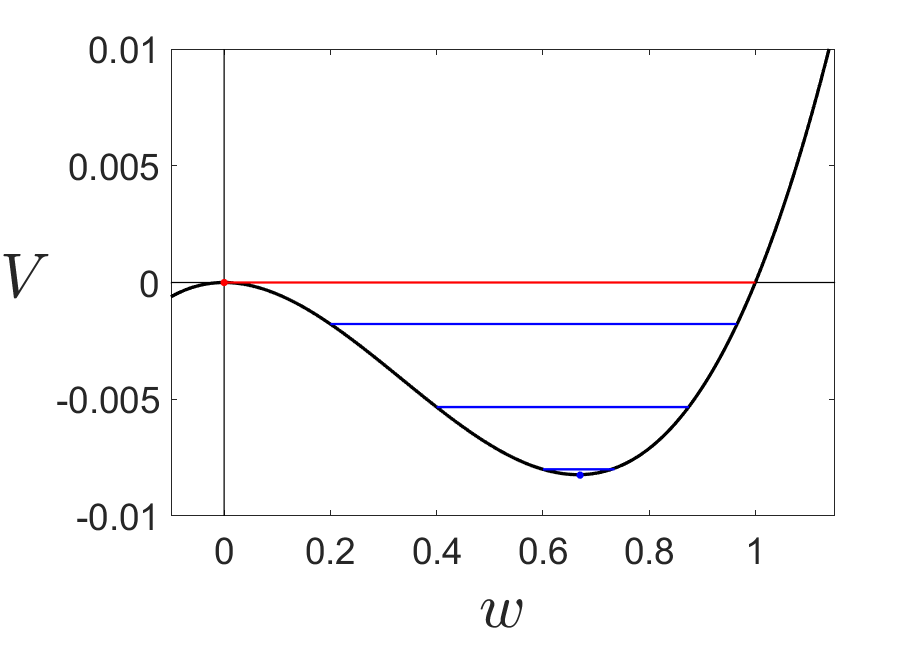}\hspace{0.5cm}
\includegraphics[scale=0.3]{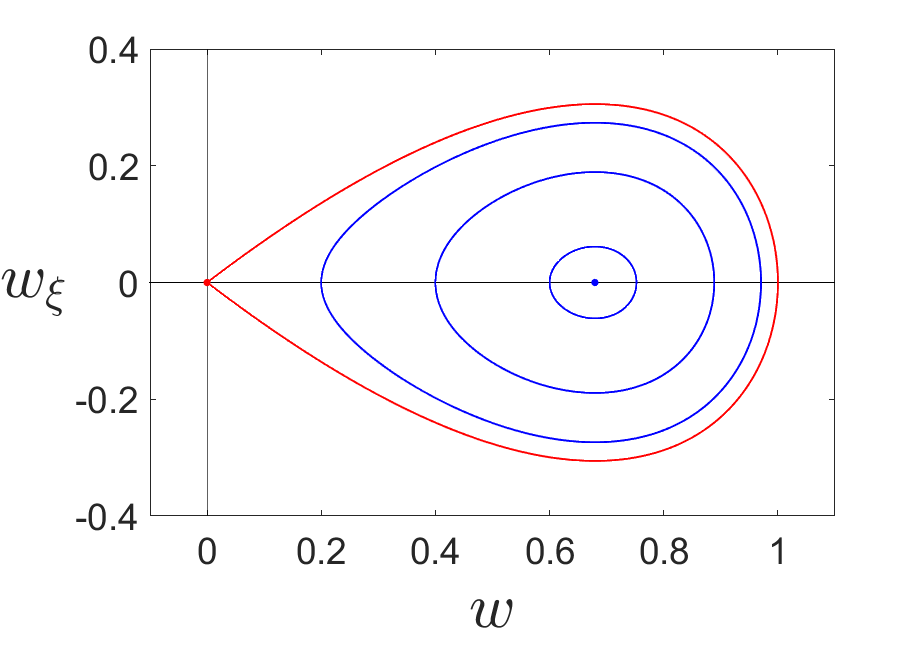}
\caption{(Color online) Left panel: The potential function \eqref{eq:V} of system with some characteristic energy levels. The colors correspond to the relative orbits in the phase-plane of the system. Right panel: The zero energy orbit corresponds to a homoclinic orbit (red line), while the negative energy orbits correspond to periodic orbits
(blue lines).}
\label{fig:V}
\end{center}
\end{figure}

In the left panel of Fig.~\ref{fig:V}, shown is the potential \eqref{eq:V}, 
for $v=1.5$ (resulting in 
$w_{min}=0.67945$), 
together with some characteristic energy  levels. The corresponding phase-portrait of the system is depicted in the right panel of the figure. As we can see, the negative energy orbits correspond to closed (periodic) orbits which correspond cnoidal wave solutions of the system. The  homoclinic orbit, for zero energy, corrsponds to a solitary wave  solution, which is not relevant to our current study. Nevertheless, it should be pointed out that this
homoclinic orbit was at the heart of the original
traveling wave concept of the Adlam-Allen study~\cite{Allen1958} (see also~\cite{Allen}).

 Since the potential is asymmetric, the oscillations around the minimum are also asymmetric. This fact becomes more obvious as we move further away from $w_{min}$, as it can be seen in Fig~\ref{fig:V}. 
 Actually, when the oscillation occurs close to the minimum $w_{min}$, it can be well approximated by a sinusoidal waveform, while as the amplitude of the oscillation increases, so do the asymmetric characteristics of the cnoidal oscillation, and Jacobi elliptic functions are required for a proper description of the corresponding waveform.

Finally, we note that, upon using \eqref{eq:dsu}, we can calculate the $u$-component of the solution, as well as the corresponding $u_{min}$ which corresponds to the potential minimum. As we will see below, the values $u_{min}, w_{min}$
are important as they suggest  the background for the multiscale analysis that we perform in the next Section.

\section{Effective NLS and dark solitons}

\subsection{Multiple scale analysis}

Having established that for energies sufficiently close to the energy correponding to the minimum of the potential, one expects the system to possess a periodic solution, we proceed as follows. 
We introduce the state vector $\textbf{S}=(u,w)$ describing  the state of the system at a given position $x$ and time $t$. 
%
%
Then, we may adopt a multiple scales perturbation technique (see details for this method in Refs.~ \cite{Jeffrey,taniuti1969,asano1969}), and proceed as follows. 
We introduce new independent spatial and temporal variables, $x_m = \epsilon^m x$, $t_m = \epsilon^m t$, where $0<\epsilon\ll 1$ is a formal small parameter, and $m=0,1,2,\ldots$. Accordingly, we expand
the space and time derivative operators as $\partial_x = \partial_{x_0} + \epsilon \partial_{x_1} + \ldots$, and $\partial_t = \partial_{t_0} + \epsilon \partial_{t_1} + \ldots$. Then, assuming a moderate (but finite) deviation from the state vector $(u_0,w_0)=(u_{min}, w_{min})$ at the local minimum of the potential $V$ of Eq.~\eqref{eq:V}, we introduce 
the following asymptotic representation of the state vector:
\begin{eqnarray}
\textbf{S} = (u_0, w_0)+ \epsilon \textbf{S}^{(1)}+ \epsilon^2 \textbf{S}^{(2)}+\ldots, 
\label{eq:ansatz01}
\end{eqnarray}
and subsequently consider multiharmonic analysis at each order.
In particular, we assume that at the order $O(\epsilon^n)$ the perturbed state $\textbf{S}^{(n)}$ can be expressed as:
\begin{equation}
\textbf{S}^{(n)} =\sum_{\ell =-\infty}^{+\infty}  \textbf{S}_{n\ell}(x_{m \geq 1}, t_{m \geq 1}) \, e^{i \ell (kx-\omega t)}.
\label{eq:ansatz02}
\end{equation}
where the subscript $\ell$ denotes the  phase-multiple 
for a given harmonic, up to the order $n$ in the expansion in $\epsilon$. More specifically, the perturbed state depends on the fast variables $x, t$ via the fundamental carrier's 
phase $kx-\omega t$ only (where $k$ and $\omega$ are the frequency and
wavenumber of the carrier wave). Furthermore, the harmonic amplitudes
$\textbf{S}_{n\ell}(x_{m \geq 1}, t_{m \geq 1})$ are slowly
varying, as they depend on the slow scales $x_m, t_m$.
%
%
%

Substituting  Eqs. (\ref{eq:ansatz01})-(\ref{eq:ansatz02}) into Eqs. (\ref{eq:eq1})-(\ref{eq:eq2}) and collecting the terms arising in each  order in $\epsilon$, we obtain the amplitude evolution equations at successive orders. In particular, up to the order $O(\epsilon^2)$, we derive the following results.
%


At order
$O(\epsilon^0)$, 
we obtain the following algebraic equation connecting $u_0$ and $v_0$:
\begin{equation}
u_0(1+w_0)=-w_0.
\label{eq:dc}
\end{equation}
It is convenient to express $u_0$ and $v_0$ in terms of an auxilliary parameter $\mu=1+w_0>1$. This way, we have: 
\begin{eqnarray}
w_0=\mu -1, \quad u_0=\frac{1-\mu}{\mu}.
\end{eqnarray}
Note that for $\mu=1$ leads to  $(u_0, w_0)=(0,0)$ .

At order $O(\epsilon)$ for the zeroth harmonic ($\ell=0$), we obtain $u_{10}=-w_{10}=0$, while for the first harmonic ($\ell=1$), we obtain the following system for $u_{11}$ and $w_{11}$:
\begin{eqnarray}
&&-\omega^2u_{11} -\mu k^2w_{11}=0,
\label{eq:eq11} \\
&&-\mu u_{11}-\left(\frac{1}{\mu}+k^2\right)w_{11}=0.
\label{eq:eq21}
\end{eqnarray}
The above system possesses nontrivial solutions as long as the determinant of the coefficients vanishes. This leads to the dispersion relation:
\begin{equation}
\omega^2=\frac{\mu^3}{1+\mu k^2}k^2.
\label{eq:linear_disp}
\end{equation}
Furthermore, employing the dispersion relation, we obtain from the system~(\ref{eq:eq11})-(\ref{eq:eq21}) the following algebraic equation connecting $u_{11}$ and $w_{11}$:
\begin{eqnarray}
&&u_{11}=-\frac{1}{\mu^2-\omega^2} w_{11}.
\label{eq:u}
\end{eqnarray}

Next, at order  $O(\epsilon^2)$, for the zeroth harmonic ($\ell=0$), we have:
\begin{eqnarray}
&&u_{20}=F_{20}|w_{11}|^2, 
\quad 
~F_{20}:= 
 \frac{2}{\mu(\mu^2-\omega^2)}-\frac{1}{\mu^2}\left[\mu \frac{\mu^4+2(\mu^2-\omega^2)^2}{(\mu^2-\omega^2)^3-\mu^6}\right],\\
&&w_{20}=G_{20}|w_{11}|^2, \quad 
G_{20}:=
\mu\frac{\mu^4+2(\mu^2-\omega^2)^2}{(\mu^2-\omega^2)^3-\mu^6}.
\label{eq:zero_harm2}
\end{eqnarray}
In addition, demanding for the secular terms 
appearing in the equations for the first harmonic ($\ell=1$) that they be zero, we obtain 
 $\partial w_{11} /\partial t_1 + v_g  \partial w_{11} /\partial x_1 = 0$, where $v_g$ is the group velocity given by:
\begin{equation}
v_{g}=
\left(\frac{\mu^2-\omega^2}{\mu}\right)^{3/2}.
\label{eq:gvelqc}
\end{equation}
The above result is consistent with the  definition 
$v_{g}:= \partial \omega(k)/\partial k$, with $\omega(k)$ given by Eq.~(\ref{eq:linear_disp}). Furthermore, at the same order, $O(\epsilon^2)$, we obtain the following equations for the second harmonic ($\ell=2$):
\begin{eqnarray}
&&u_{22}=F_{22}w_{11}^2, \quad 
~F_{22}:=
-\frac{\mu^2+\omega^2}{2\omega^2\mu(\mu^2-\omega^2)} 
\label{eq:sec_harm21} \\
&&w_{22}=G_{22} w_{11}^2, \quad 
G_{22}:=\frac{\mu}{2\omega^2}w_{11}^2.
\label{eq:sec_harm22}
\end{eqnarray}

Finally, at order $O(\epsilon^3)$, the requirement that the secular terms arising in the equations for the first harmonic  ($\ell=1$) be zero, 
leads (as is typical in such settings~\cite{Jeffrey}), to the following NLS equation:
\begin{equation}
 i \frac{\partial \Psi}{\partial \tau} + P \frac{\partial^2 \Psi}{\partial \xi^2}  + Q |\Psi|^2 \Psi=0.
\label{eq:NLSE} \\
\end{equation}
Here, $\Psi:=w_{11}$,  
the (slow) time and space variables are defined as:
\beq
\tau = t_2, \quad \xi = x_1 - v_g t_1,
\label{eq:scaling}
\eeq
while the dispersion coefficient $P$ and nonlinearity [self-phase modulation (SPM)] coefficient $Q$ are  respectively given by: 
\begin{eqnarray}
 P &\equiv&\frac{1}{2}\frac{\partial^{2}\omega}{\partial k^{2}}=-
\frac{3\omega({\mu^2-\omega^2})^2}{2\mu^3},
\label{eq:dispcoef}  \\
 Q&=&-\frac{\omega (\mu^2-\omega^2)}{2\mu}\left[\frac{2}{\mu^2-\omega^2}\left(G_{20}+G_{22}\right)-\left(F_{20}+F_{22}\right)\right] \nonumber \\
&=&\frac{9\mu^6-12\mu^4\w^2+8\mu^2\w^4-\w^6}{4\mu^2(3\mu^4\w-3\mu^2\w^3+\w^5)}.
\label{eq:nlcoef1}
\end{eqnarray}
The coefficients $P$ and $Q$ are essentially real functions of the carrier wavenumber $k$ [recall Eqs.~(\ref{eq:linear_disp}) and (\ref{eq:gvelqc}) above]; 
the frequency depends on $k$, and $\mu$ is fixed by
the point around which we are expanding.  

It should also pointed out that, generally, a NLS equation of the form~(\ref{eq:NLSE}) for $PQ<0$ the NLS model supports dark soliton solutions,
while for the case of $PQ>0$, the model possesses
bright solitary wave solutions~\cite{Kivshar1989,Kivshar2003}. Hence, our natural next step is to examine the nature 
(especially the sign) of this product and accordingly
use the resulting NLS model to approximately reconstruct
the corresponding solitary waves of the original AA model,
whose potential robustness needs to be explored numerically.

\subsection{Dark solitons}

\begin{figure}[tbp]
\begin{center}
\includegraphics[scale=0.3]{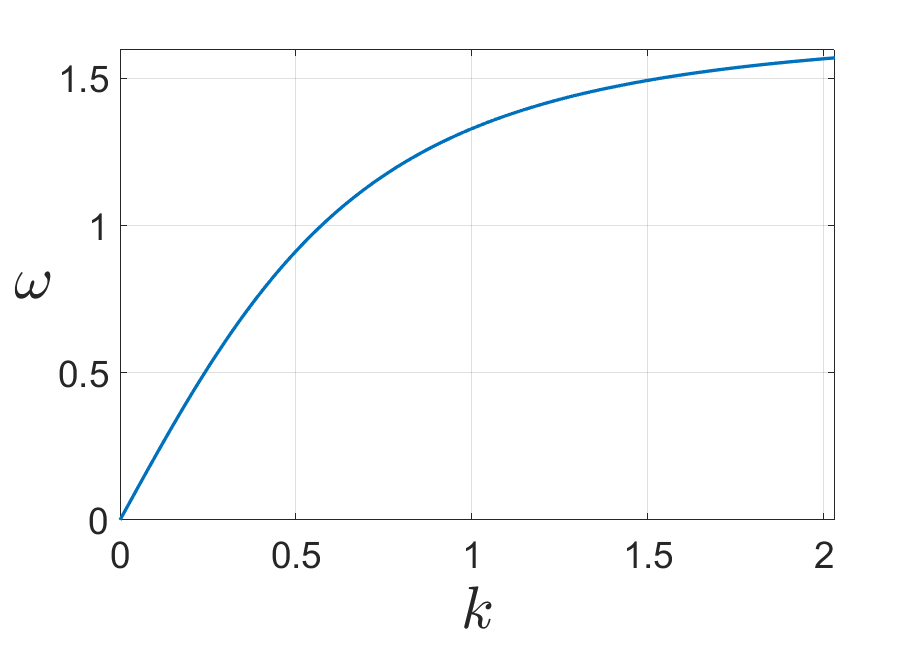}    
\includegraphics[scale=0.3]{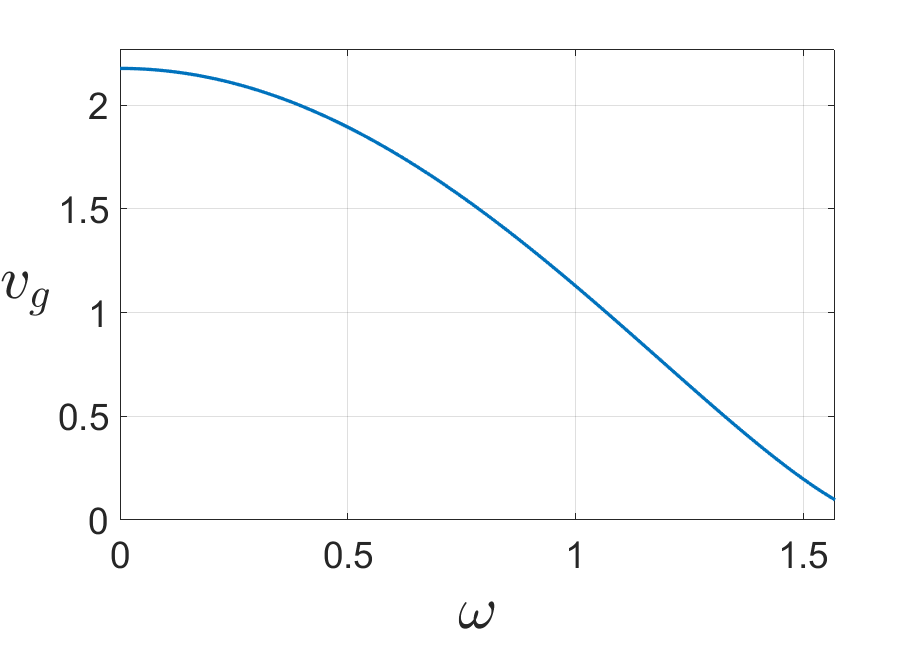}
\includegraphics[scale=0.3]{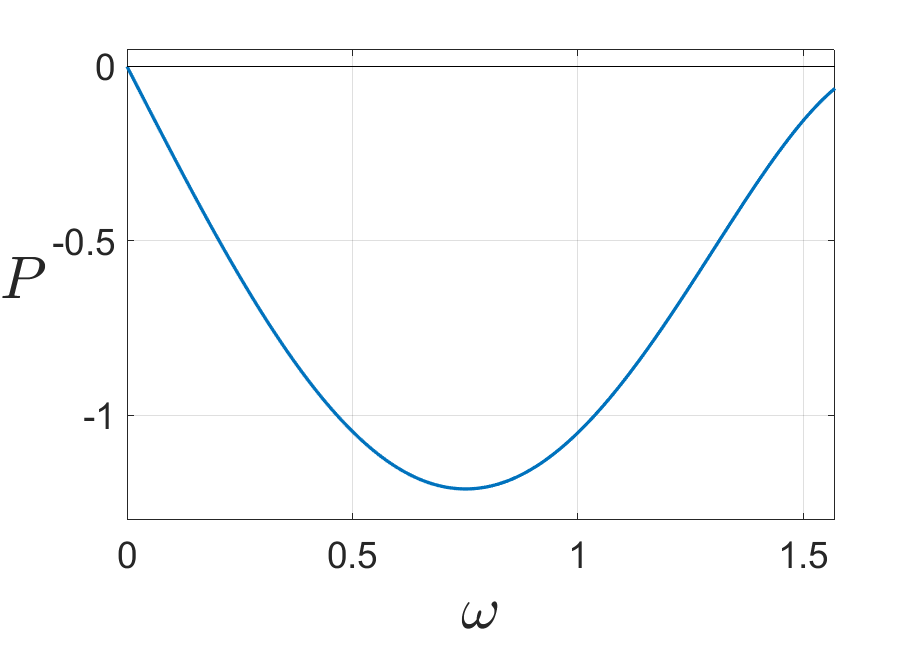}
\includegraphics[scale=0.3]{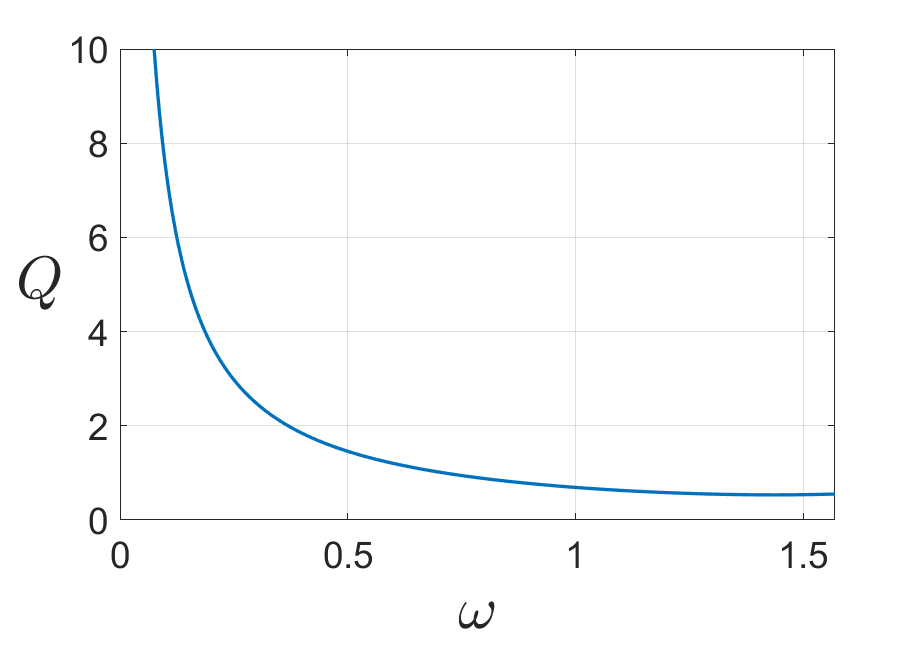}
\end{center}
\caption{(Color online) Top panels: The dispersion relation (left) $\omega=\omega(k)$,  
for $\mu=1.67945$. Right panel: 
the group velocity $v_g$ (right) as 
a function of the frequency $\omega$ for $\mu=1.67945$.
Bottom panels: The dispersion coefficient $P$ (left) and the nonlinearity coefficient $Q$ (right) as functions of 
the frequency $\w$ 
the same value of $\mu$. Their product is negative for every value of the frequency.
}
\label{fig:dispersion}
\end{figure}

We consider a value for $w_0$ in the interval $0<w_0<1$. In particular, for our investigation, our choice is $w_0=0.67945$, corresponding to the minimum of the effective potential (\ref{eq:V}) for $C=1.5$, as explained above; as a result, 
the parameter $\mu$ takes the value $\mu=1.67945$. For this choice, in the top panels of Fig.~\ref{fig:dispersion}, we show the dispersion relation \eqref{eq:linear_disp} and the group velocity $v_g$ with respect to the frequency $\w$ [see Eq.~(\ref{eq:gvelqc})]. 
On the other hand, the dispersion coefficient $P$ [Eq.~\eqref{eq:dispcoef}] and the nonlinear coefficient $Q$ [Eq.~\eqref{eq:nlcoef1}] are depicted, as functions of $\omega$, in the bottom panels of the same figure. 
It is evident that the product of the two coefficients is negative. 
Here, we have to underline that, although we depict $P$ and $Q$ only for the specific choice of $\mu$, their product remains negative for all values of $\mu$ in the permitted parameter region $1<\mu<2$ (since $0<w<1$, as illustrated, e.g., in Fig.~1 of~\cite{Allen}).

According to the above, for $PQ<0$, it is expected that the NLS~\eqref{eq:NLSE}, supports dark soliton solutions. The single dark soliton of \eqref{eq:NLSE}, in its most general form, is given by~(see Ref.\cite{DJF}):
\begin{equation}
\Psi(\xi,\tau) = \sqrt{\eta_0} {\{B_s\rm tanh}\left[\sqrt{\eta_0}B_s(\xi-U_{s}\tau-\xi_0)\right]+iA_s\}e^{i(K_s\xi-\Omega_s \tau+\thet_0)}.
\label{sol1}
\end{equation}
Here, $\sqrt{\eta_0}$ is a $O(1)$ parameter connected with the soliton's amplitude,
$U_s=A_s\sqrt{n_0}+K_s$ is the  relative velocity between the soliton and the background, $K_s$ and  $\Omega_s=(1/2)K_s^2+\eta_0$ are the soliton's wavenumber and frequency respectively, 
and the parameters $A_s$ (connected with the soliton velocity), and $B_s$ (connected with the soliton depth) are connected through the equation $A_s^2+B_s^2=1$. Finally, 
$\xi_0$ and $\thet_0$
represent, respectively, the initial location and the phase of the dark soliton. 

In many physically relevant situations, the background is at rest, i.e., $K_s=0$. In this case, and assuming for simplicity that $\xi_0=0$, $\thet_0=0$, we may employ \eqref{eq:scaling} and express the solution~(\ref{sol1}) in terms of the variables $x$ and $t$ as follows:
%
%
\beq
\Psi(x,t) = \sqrt{2\eta_0\left|\frac{P}{Q}\right|}\left\{B_s\tanh\left[\e B_s\sqrt{\eta_0}\left(x-(v_g+2\e\sqrt{\eta_0}PA_s)t\right)\right]
+iA_s\right\}e^{-i2\eta_0P\e^2t}.
\label{sol2}
\eeq
%
Thus, the dark soliton solution of the AA model~(\ref{eq:eq1})-(\ref{eq:eq2}) estimated by the perturbation theory is of the form:
\begin{eqnarray} w(x,t)&=&w_0+\e 
[\Psi(x,t)e^{i(kx-\w t)}+{\rm c.c.}]+\cal{R},
\label{eq:sol3}
\\
u(x,t)&=&u_0-\frac{1}{\mu^2-\omega^2}\e 
[\Psi(x,t)e^{i(kx-\w t)}+{\rm c.c.}]+\cal{R},
\label{eq:sol4}
\end{eqnarray}
where 
\begin{eqnarray}
\Psi(x,t)e^{i(kx-\w t)}+{\rm c.c.} =
\sqrt{8\eta_0 \left|\frac{P}{Q}\right| }
&\Big\{&B_s\tanh\left[\e\sqrt{\eta_0}B_s(x-(v_g+2\e PA_s\sqrt{\eta_0})t\right]
\cos\left[kx-(\omega+2\epsilon^2 P\eta_0)t\right]
\nonumber \\
&-&A_s\sin\left[kx-(\omega+2\e^2 P\eta_0)t
\right]
\Big\},
\label{eq:darksolution}
\end{eqnarray}
and 
%
the remainder ${\cal R}(\textbf{S}_{(n \ell )} )$ with $|\ell|>1$ for $n=1$ and $\ell\in\mathbb{Z}$ for $n>1$ can be neglected. 
%

Finally, by also using \eqref{eq:RB2uw}, we acquire the initial conditions $R(x,t_0)$ and $B(x,t_0)$ which  are used as an input to an appropriate numerical scheme for the integration of the original system \eqref{eq:aa1}-\eqref{eq:aa2}.
Below, we will  consider only the black soliton case ($B_s=1$, $A_s=0$), since the grey soliton ($A_s \neq 0$) features a similar 
dynamical behavior. 

\section{Numerical study}

\subsection{The integration scheme}

Our aim is to numerically integrate the original AA system \eqref{eq:aa1}-\eqref{eq:aa2}, which consists of a wave-like equation [Eq.~\eqref{eq:aa1}] and
a Poisson-like equation [Eq.~\eqref{eq:aa2}]. The second one does not possess an evolution component. Thus, the time-evolution of the system, as a whole, cannot be calculated by integrating both equations simultaneously. In order to tackle this problem we used a two-step approach.

First, we set up the initial conditions $R(x,t_0), B(x,t_0)$, which are obtained as described in the previous Section, and we use them to advance the first equation by one time-step $\Delta t$. Then, we use the advanced value of $R^{(1)}=R(x,t_0+\Delta t)$ to the second equation and solve it with respect to $B$ to obtain $B^{(1)}$, which is the value of $B$ which is compatible with $R^{(1)}$.

For the time integration of the system, an explicit constant step Runge-Kutta of the $8^{th}$ order has been used, 
and has been found to be sufficient for our purposes. 
%
%
As for the spatial discretization, an $8^{th}$-order finite-difference formula was used for the second derivative. In order to confirm the validity of our findings, a Fourier-spectral scheme was also used, providing completely similar results. The finite-differences method 
was finally preferred due to its favorable performance in terms of integration time. In addition, we considered periodic boundary conditions;  the space domain was always chosen to be sufficiently large to avoid boundary effects, although, in the numerical results presented in the next section, usually we depict only the part of the space-domain which is relevant to our study. 

\subsection{Numerical results}

\begin{figure}[tbp]
\begin{center}
\includegraphics[scale=0.28]{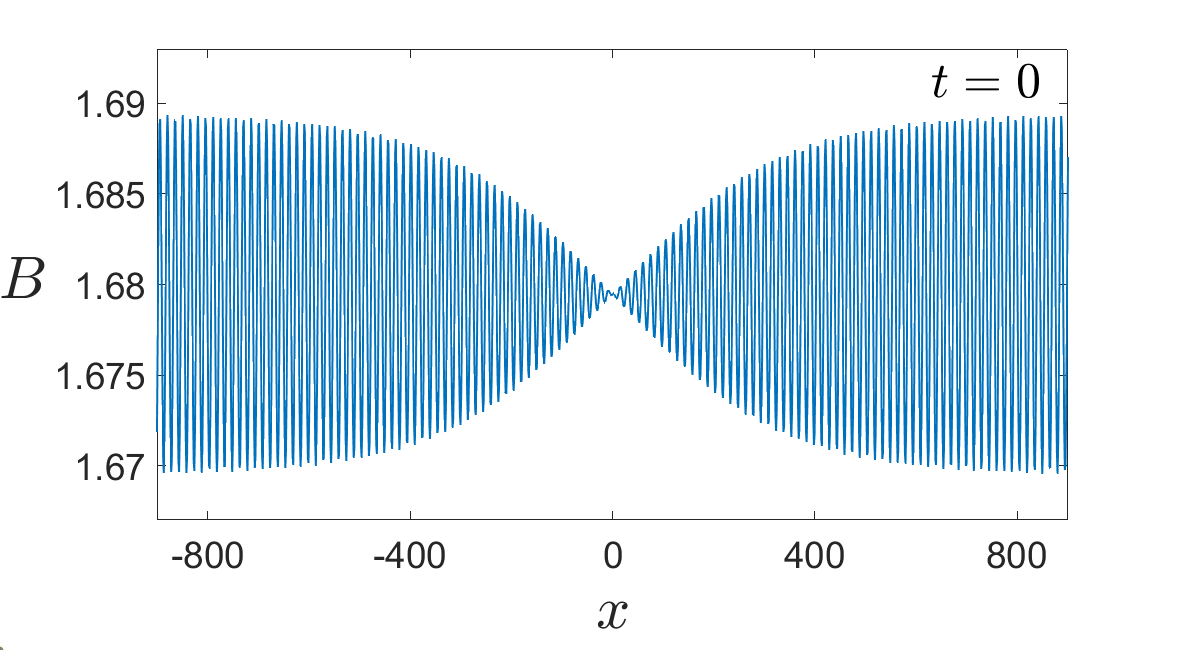}
\includegraphics[scale=0.28]{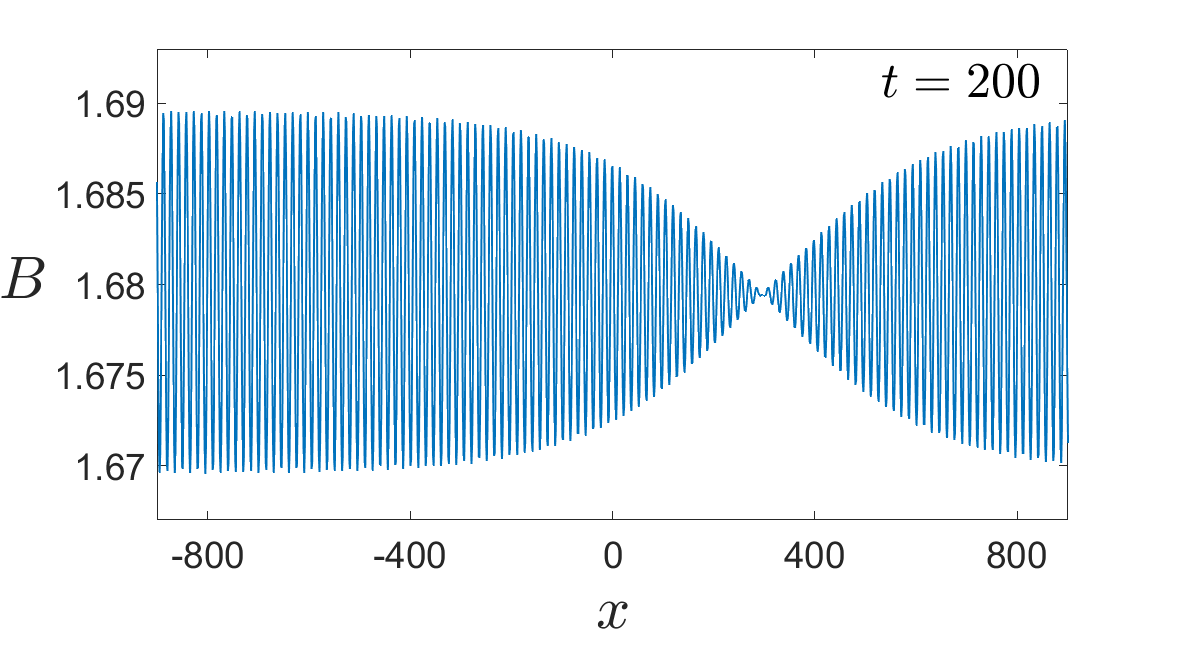}\\
\includegraphics[scale=0.28]{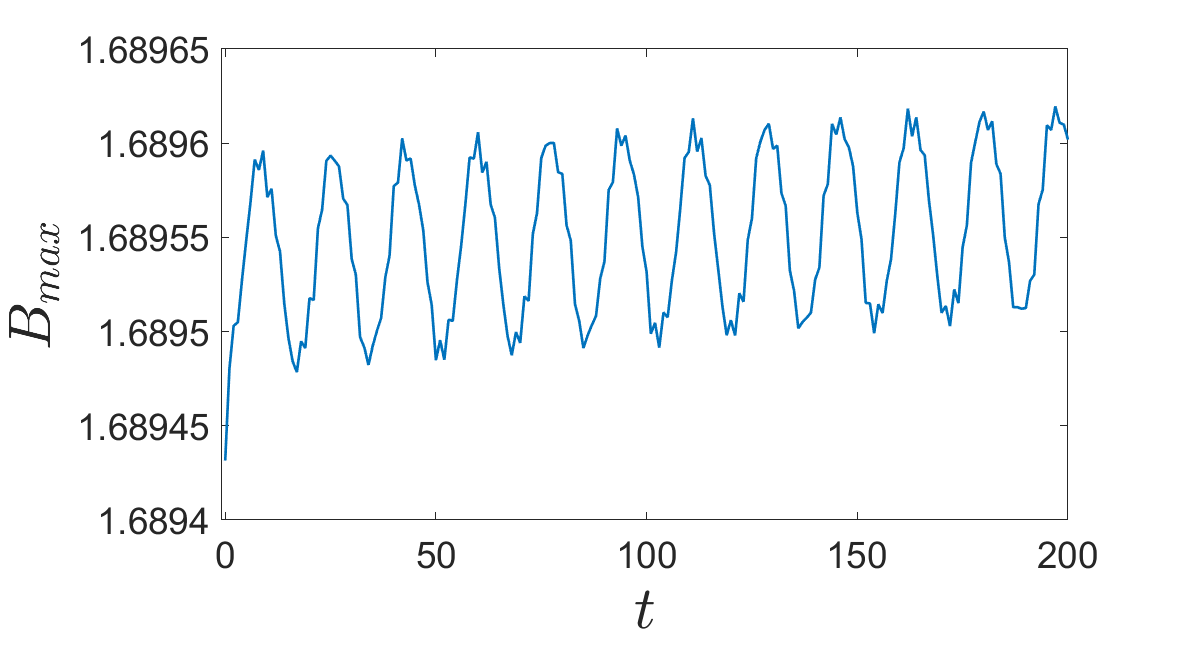}
\includegraphics[scale=0.28]{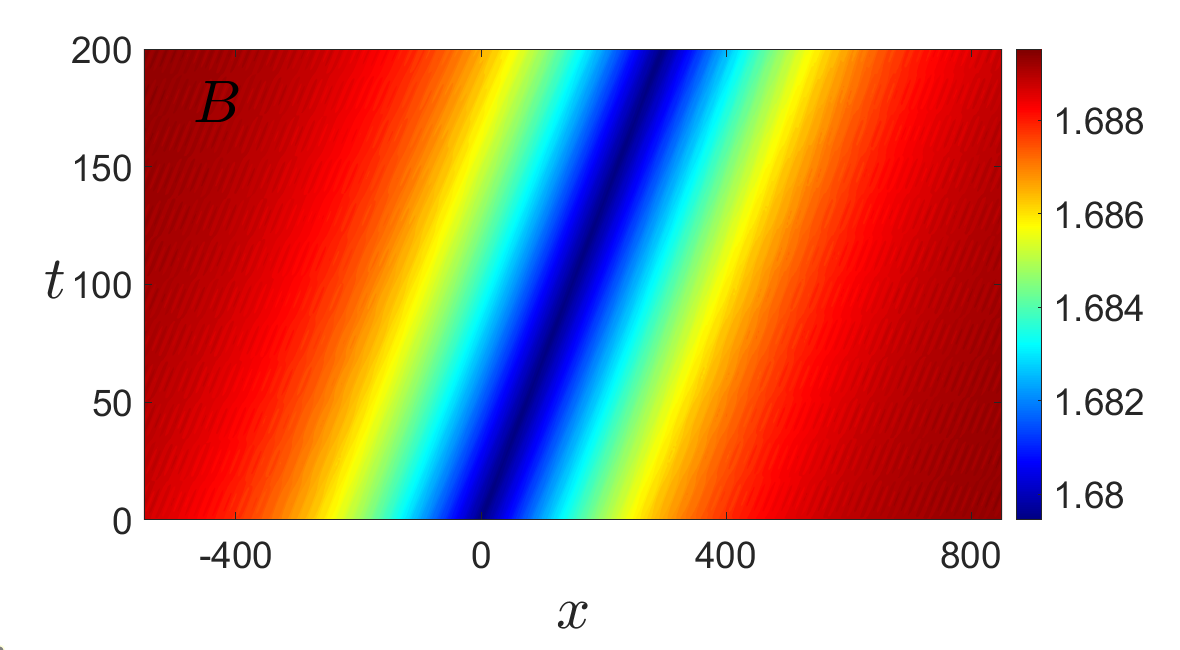}
\caption{(Color online) The small amplitude case ($\e=0.01$). Upper panels: two snapshots of $B(x,t)$ for $t=0$ and $t=200$. Lower left panel: the temporal variation of the maximum of $B$. Lower right panel: a contour plot of the space-time evolution of $B$, showcasing the traveling wave nature of the relevant (dark solitonic) dip in the magnetic field.}
\label{fig:e_0_01}    
\end{center}
\end{figure}

The initial conditions for the numerical study of this section are taken by inserting \eqref{eq:darksolution} into the perturbation-theoretic expressions
of Eqs.~\eqref{eq:sol3}-\eqref{eq:sol4} evaluated at $t=0$. We first consider the case of a small-amplitude dark soliton. Since the solution amplitude is analogous to $\e$, we set the 
formal small parameter to $\e=0.01$, while we set the free parameter $\eta_0$ in \eqref{eq:darksolution} to be $\eta_0=|Q/8P|$ (so that the factor $\sqrt{8\eta_0|P/Q|}=1$). 
This choice allows us to directly connect the soliton amplitude with the value of $\e$. The results of this investigation are depicted in Fig.~\ref{fig:e_0_01}.
In the upper panels of Fig.~\ref{fig:e_0_01} we see the initial state ($t=0$) of the magnetic field $B$ and the corresponding one for $t=200$. The inverse density $R$ is not depicted here for the sake of a more succinct presentation, since it does not provide any additional information about the dynamics of the system (i.e.,
it is fairly similar to the $B$ dynamics shown).
In these panels we can see that, for this choice of $\e$, the perturbation theory provides an excellent approximation of the actual solution. The soliton propagates practically intact. There is a slight oscillation of the background of the solitary wave which is practically unseen (to the naked eye). In the lower left panel of Fig.~\ref{fig:e_0_01} this oscillation is measured by calculating the temporal variation of the maximum value of $B$. This is found to be of $O(\e^2)$ as is expected on the basis of the expansion, since our approach provides $O(\e)$ approximations of the exact solution. It is worthwhile to mention that in the first time-step we observe a ``jump'' at the value of $B_{max}$. 
This is a product 
%
of the (partial) incompatibility of our initial conditions with the algebraic equation of interest. 
 More specifically, 
after calculating $R^{(1)}$ by integrating \eqref{eq:aa1}, we solve \eqref{eq:aa2} to find the compatible $B^{(1)}$. However,  our initial conditions are accurate only up to $O(\e)$, thus the field $B$, in order to adjust to the precision of $R$, needs to perform this $O(\e)$  
initial jump which is present in all the cases we examined in this work.
In the bottom right panel of the figure a contour plot of the space-time evolution of the magnetic field 
$B$ is provided. There we can see that the solitary wave moves practically intact at a constant velocity which is in an excellent agreement with $v_g$  given by \eqref{eq:gvelqc}.

\begin{figure}[tbp]
\begin{center}
\includegraphics[scale=0.28]{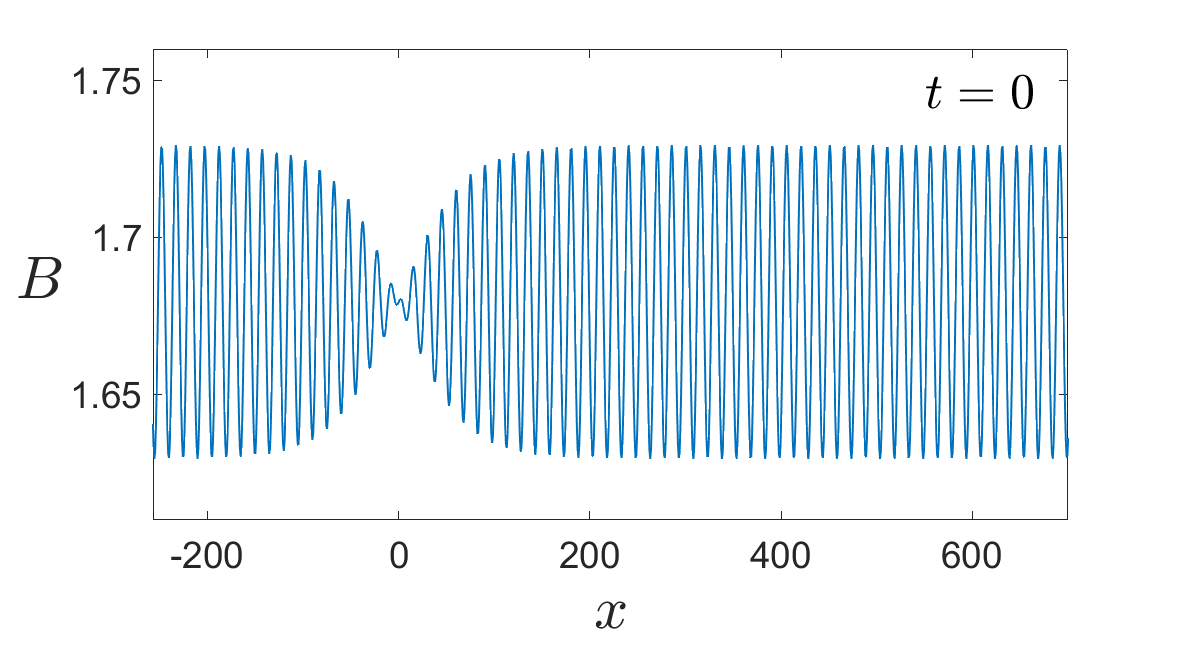}
\includegraphics[scale=0.28]{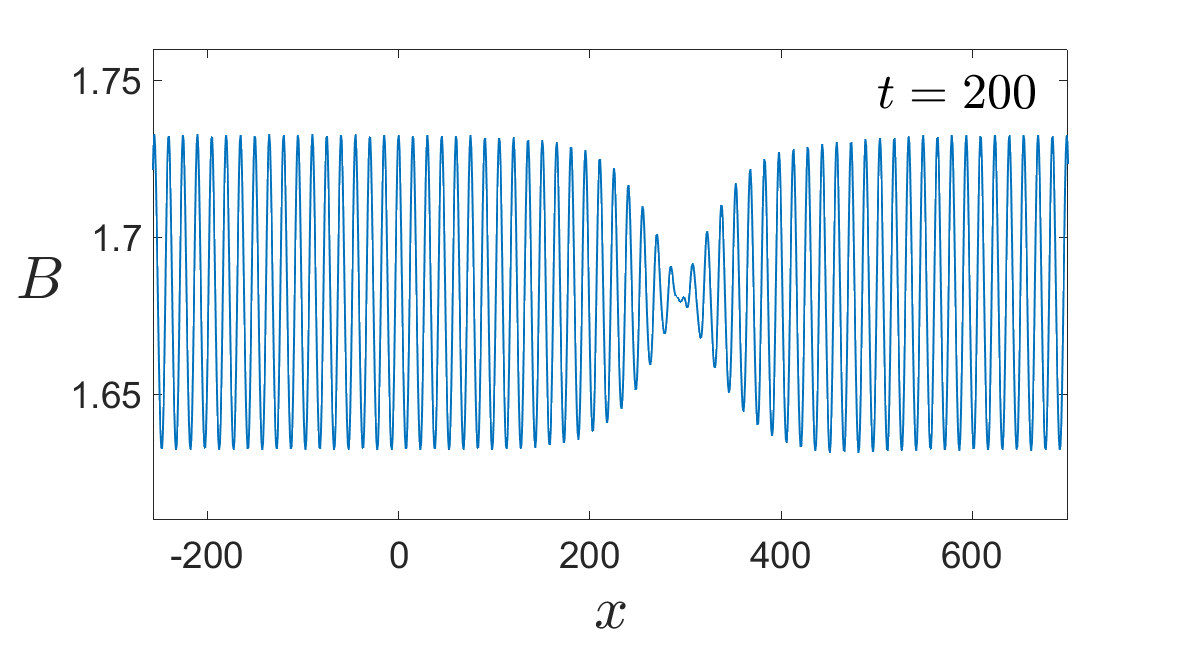}\\
\includegraphics[scale=0.28]{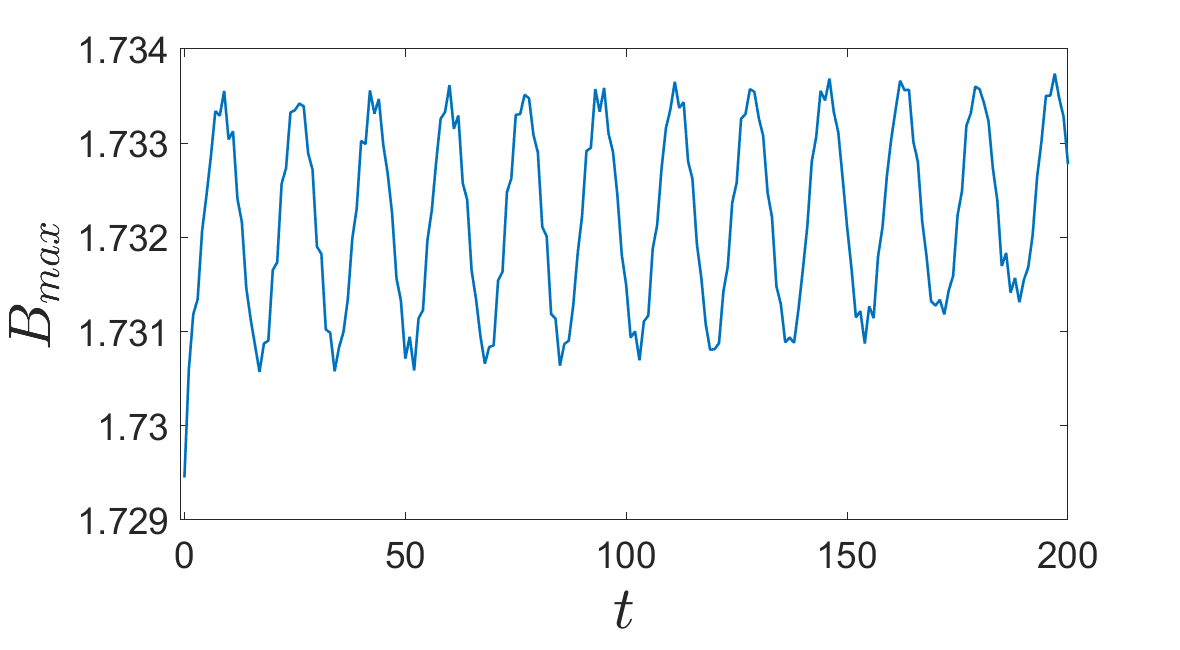}
\includegraphics[scale=0.28]{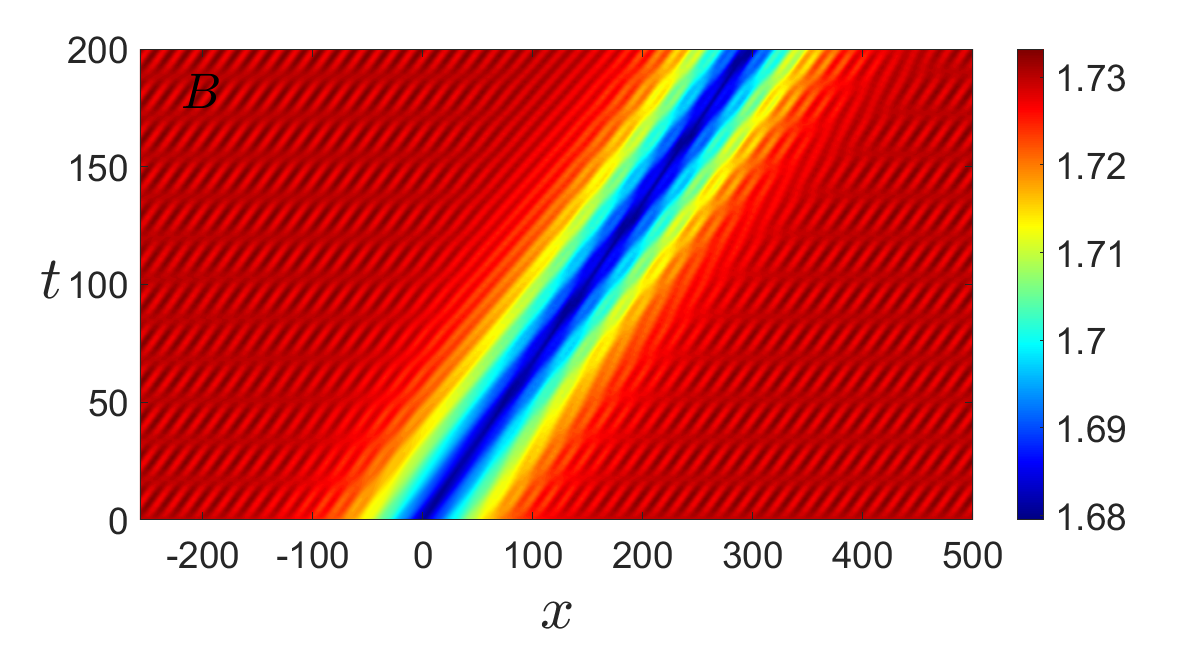}
\caption{(Color online) The case of $\e=0.05$. Upper panels: two snapshots of $B(x,t)$ for $t=0$ and $t=200$. Lower left panel: the temporal variation of the maximum of $B$. Lower right
panel: a contour plot of the space-time evolution of 
the magnetic field $B$.}
\label{fig:e_0_05}    
\end{center}
\end{figure}

After this initial evolution example, solidifying our
expectation about the dynamical relevance and structural
robustness of the waveform of interest,
we attempt to examine the range of validity of our perturbation theory approach. Thus, we consider a larger value of the perturbation parameter, i.e.,~$\e=0.05$. The results for this case are shown in Fig.~\ref{fig:e_0_05}.  Now, as can be seen in the top two panels, the oscillation of the background of the dark solitary wave becomes more evident, as well as some distortion of the solution. But, as can be seen in the bottom left panel of the figure, again this variation is of the order of $O(\e^2)$ as expected from our approximation. In the contour plot of the space-time evolution which is shown in the lower right panel of the figure, the distortion of the solution begins to appear although it still clearly manifests a coherent structure moving at a constant speed.

\begin{figure}[tbp]
\begin{center}
\includegraphics[scale=0.28]{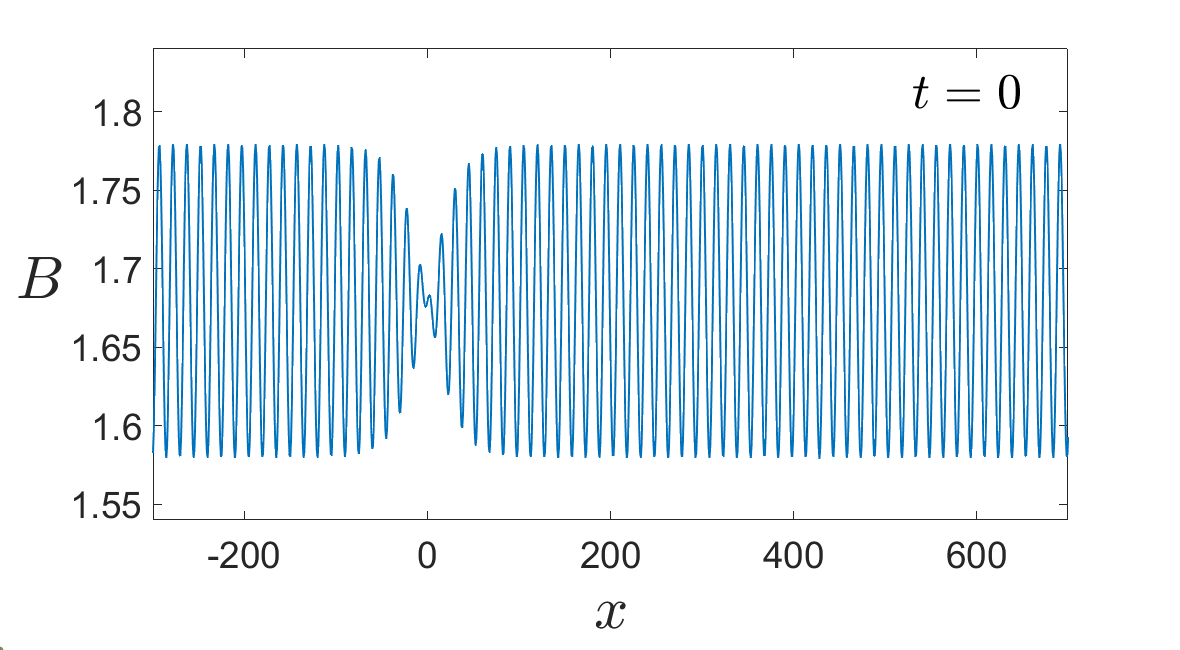}
\includegraphics[scale=0.28]{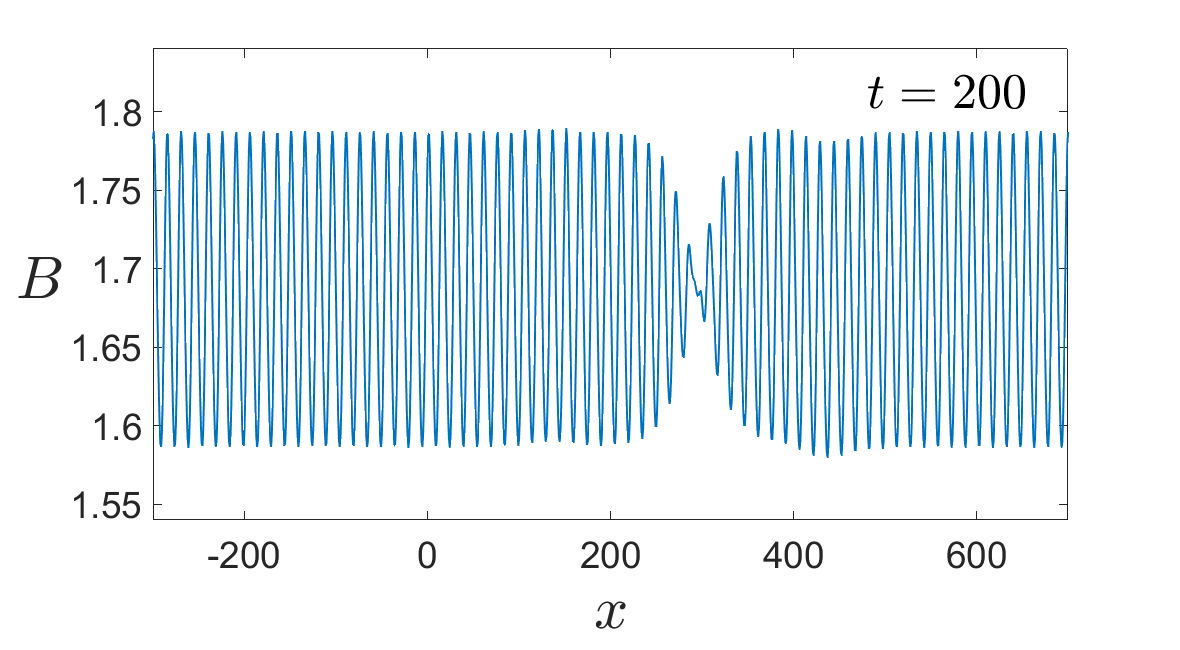}\\
\includegraphics[scale=0.28]{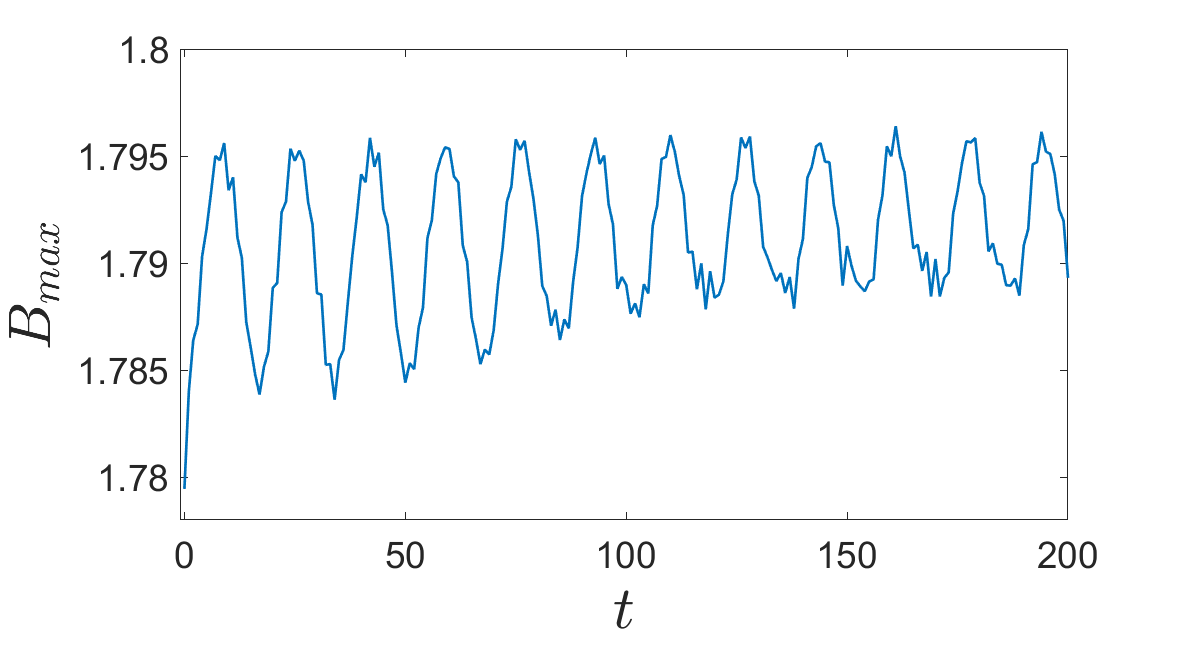}
\includegraphics[scale=0.28]{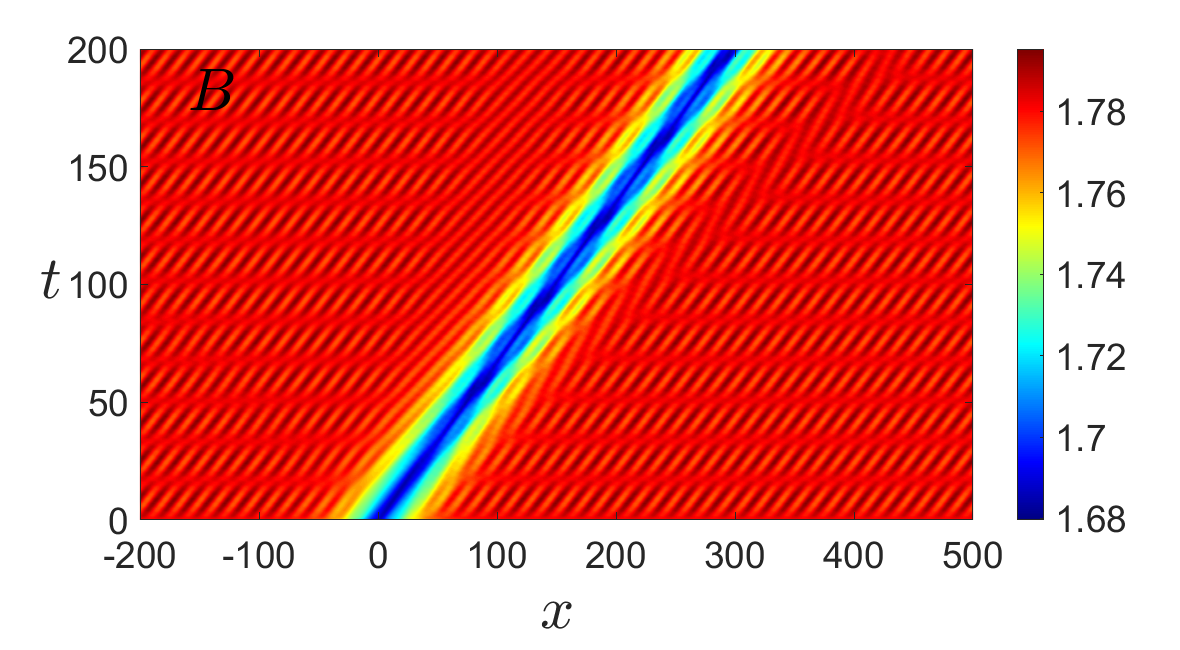}
\caption{(Color online) The large amplitude case ($\e=0.1$). 
The panels show similar features for this case as in 
the previous two figures.}
\label{fig:e_0_1}   
\end{center}
\end{figure}

Finally, we consider the case of a(n even) larger 
amplitude dark solitary wave, by taking $\e=0.1$.
%
As we can see in  Fig.~\ref{fig:e_0_1}, for this value of $\e$ the background deforms more notably acquiring a bump for $t=200$. Nevertheless, even for this value of $\e$ and for 
the time horizon considered, the temporal variation of the maximum of $B$ remains of the order of ${\cal O}(\e^2)$ which validates our approach. The gradual distortion of the solitary wave solution is also evident in the space-time evolution depicted in the bottom  right panel of the figure.

\begin{figure}[tbp]
\begin{center}
\includegraphics[scale=0.33]{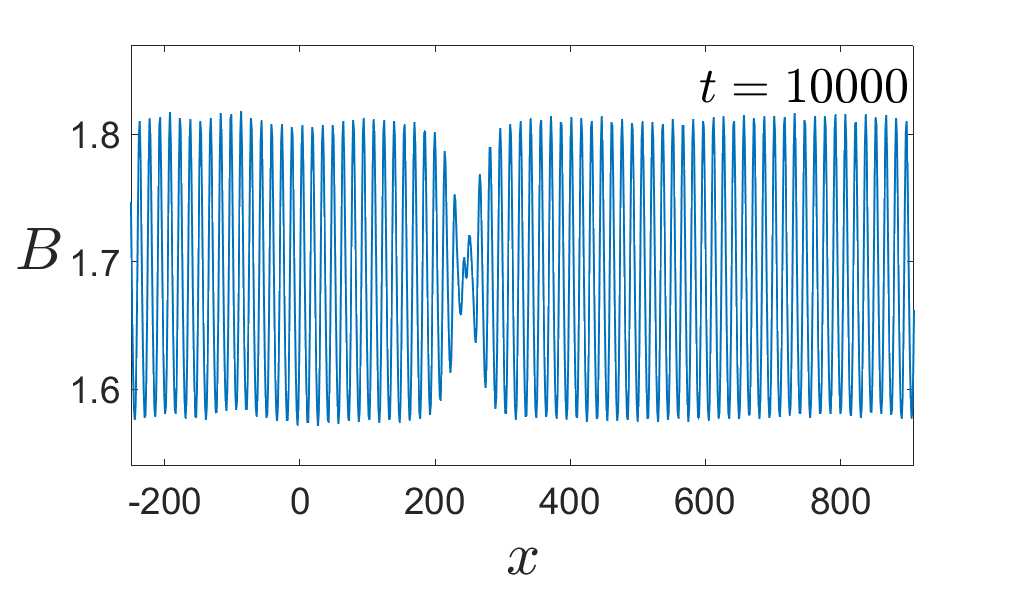}
\includegraphics[scale=0.33]{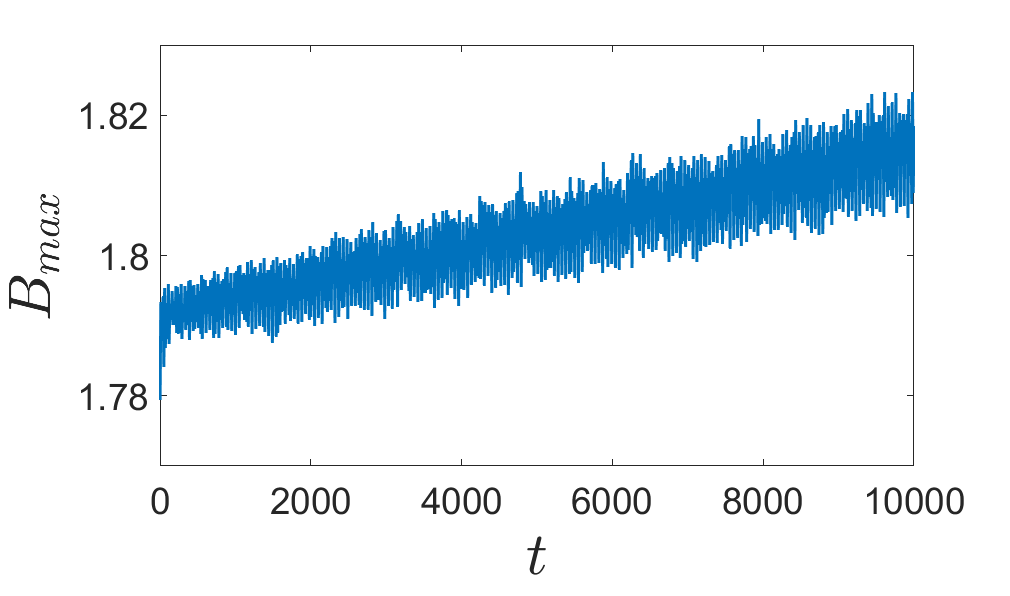}
\caption{(Color online) The long term behavior of the approximated solution for the large amplitude case ($\e=0.1$). Left panel: The background is clearly distorted but the main dark-soliton is intact. Right panel: the temporal variation of $B_{max}$ for the same time range}
\label{fig:e_0_1_long}   
\end{center}
\end{figure}

In an attempt to examine the long-term persistence of this solution, we perform a numerical integration thereof for $\e=0.1$ up to $t=10000$. As can be seen in Fig.~\ref{fig:e_0_1_long} although the background continues to oscillate and  is obviously distorted, its coherent
structure, i.e., the
dark soliton waveform remains intact. The increasing values of $B_{max}$, as well as the increasing oscillation amplitude, are caused from the accumulated numerical integration error (over such
large integration times).


In all our numerical treatment so far, we observe that the oscillation of the background of the soliton is the main cause of the deformation of our solutions. In the following Section  
we will provide an explanation for this behavior, and we will attempt to construct a more accurate dark solitary wave structure.

\subsection{A modified (tailored-background) dark solitary
wave}


Returning to our dark solitary wave structure, the background of the real soliton solution should be a cnoidal wave like the one described above. Since the oscillation of the background occurs around $w_{min}$, this justifies the choice of $w_0=w_{min}=0.67945$ as well as the choice of $\mu=1+w_0=1.67945$ as explained in Section~\ref{cnoidals}. In addition, it is reminded that $w_0$ should be chosen in the region $0<w_0<1$ and consequently $1<\mu<2$.

Since our approximate solution \eqref{eq:darksolution}
is constructed order-by-order and accordingly can only
provide a sinusoidal background (at the level of a linear
rather than a progressively more nonlinear, oscillator), it cannot capture entirely the background of the full solution. This discrepancy becomes more obvious as we consider larger values of $\e$ and thus larger amplitudes for the solitary wave. This fact causes the oscillation of the background of the reconstructed waveform that we observed in the previous section.


In an attempt to manually construct a more accurate dark solitary wave approximation, we adopt the following procedure. We  numerically calculate a cnoidal wave of $O(\e)$ amplitude;  this periodic structure is then used as 
a more accurate background, adequately incorporating the nonlinear oscillations, which
our dark solitary wave modulates. This is achieved by considering the initial condition $w(0)=w_{min}-\e, w_{\xi}(0)=0$ and numerically integrating the dynamical system \eqref{eq:dsw} for a large enough domain of $\xi$. After acquiring the background of the wave structure,
we apply an envelope modulation of the form 
$$w_{mod}=\tanh(\e\sqrt{\eta_0}(x-v_gt)),$$
where $v_g$ is the one given by \eqref{eq:gvelqc}. The specific values of $\eta_0$ and $v_g$ are the ones used in the previous section while the form of the modulation is dictated by the form of \eqref{eq:darksolution}.

\begin{figure}[H]
\begin{center}
\includegraphics[scale=0.32]{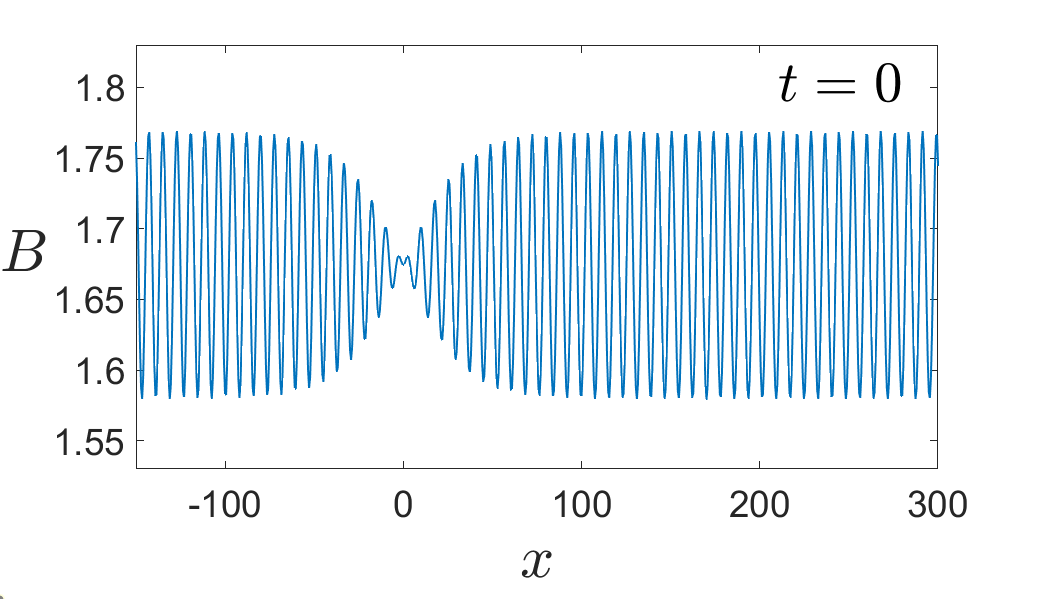}
\includegraphics[scale=0.32]{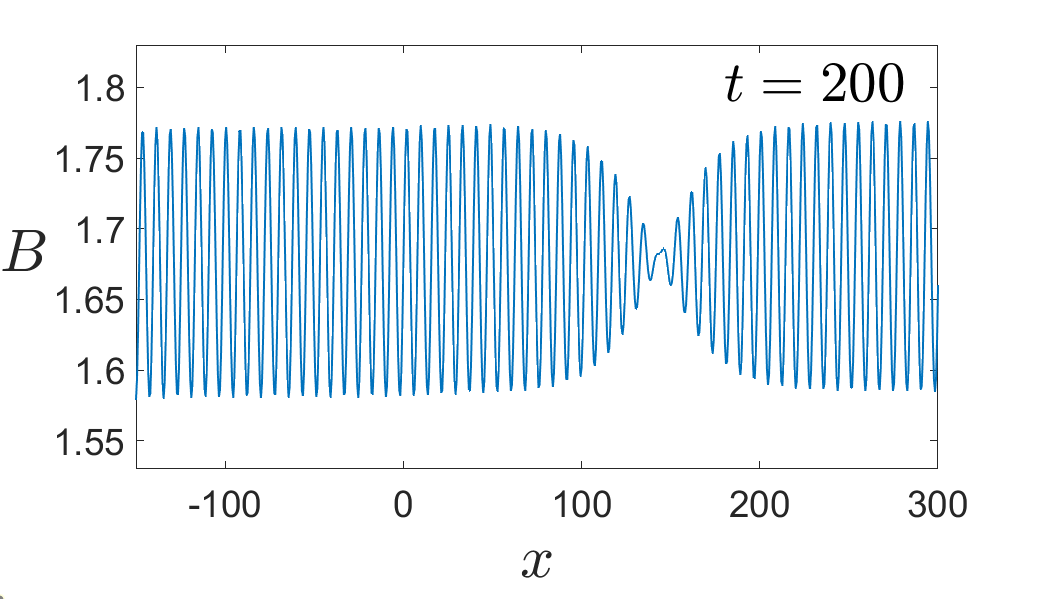}\\
\includegraphics[scale=0.32]{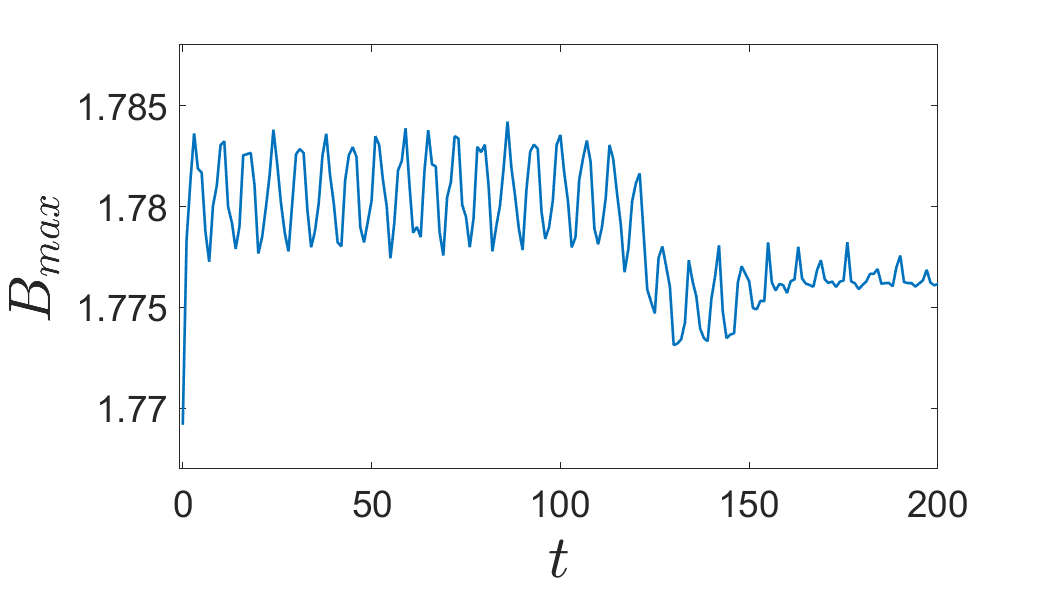}
\includegraphics[scale=0.32]{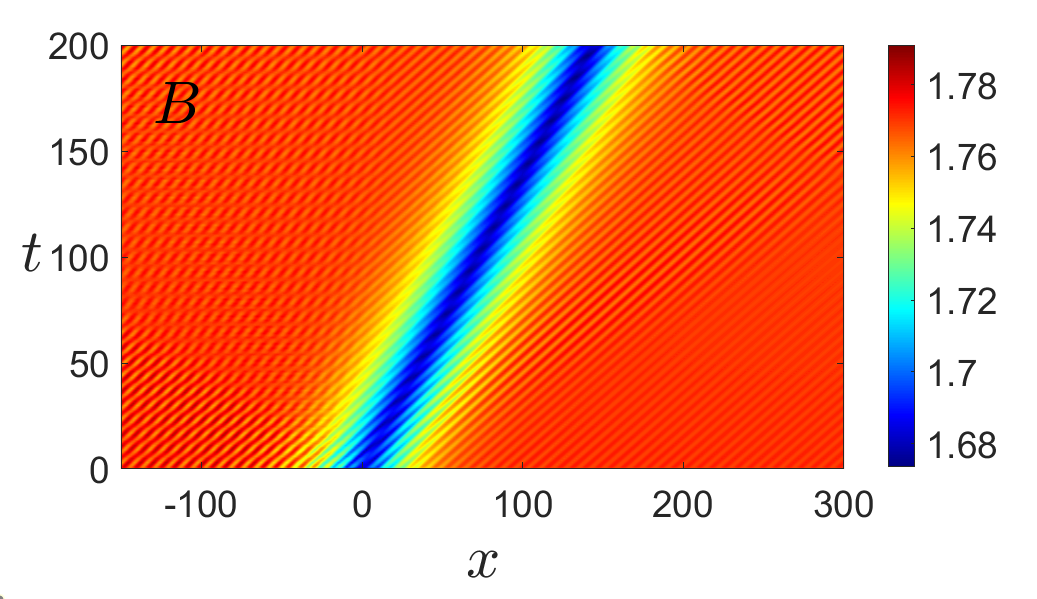}
\caption{(Color online) Constructed dark soliton solution for $\e=0.1$. Top  panels: Two snapshots of $B(x,t)$ for $t=0$ and $t=200$. Bottom-left panel: The temporal variation of the maximum of $B$. Bottom-right panel: A contour plot of the space-time evolution of $B$}
\label{fig:cnoidal_e_0_1}
\end{center}
\end{figure}

We consider here only the large amplitude case, i.e.,~ the one with $\e=0.1$. The results of this study are shown in Fig.~\ref{fig:cnoidal_e_0_1}. In the lower left panel of this figure we see that, as expected, the oscillation of the background decreases significantly. Specifically, the temporal variation of the background is half of the corresponding one in Figure~\ref{fig:e_0_1} until $t=120$, allowing for
a remarkably less distorted evolution.
After this point, the solution seems to settle to a more stable background state. Moreover, in the time evolution of the soliton which is depicted in the two upper panels we see that the soliton propagates rather intact, as is also verified by the lower right contour plot. 

\section{Conclusions and Future Challenges}

In the present work, we have revisited the classic
problem of the Adlam-Allen (AA) model of 
cold collisionless plasmas. This model was
unfortunately long forgotten as concerns
its ability to feature electromagnetic 
traveling nonlinear waves in a non-integrable
form of two reduced PDEs, arising from a system of Maxwell's and fluid equations
and its suitably tailored electric and magnetic field,
under the assumption of quasineutrality.
Here, we continued our 
studies of the 
rich features of this model. In earlier studies
we focused more on its quintessential traveling 
solitary waves and their interactions. The emphasis
of the present study was on moving away from
the saddle point (of the co-traveling frame, phase plane
analysis) that gave rise to the latter and toward
the vicinity of the center point that gives rise to 
oscillatory features and cnoidal waves.
We revealed that the vicinity of such
a center can give rise via a multiple scales
expansion at the suitable cubic order to a nonlinear
Schr{\"o}dinger effective description, which, for
realistic parameters of the AA model, turns out to be
generically defocusing. This, in turn, prompted us
to propose and successfully test the emergence of 
dark solitary waves in the AA mode, which have
never been reported before to the best of our
knowledge. While the multiscale expansion built
such structures order-by-order on top of a trigonometric
modulation background that progressively became less
adequate due to the nonlinear nature of the model's
oscillatory patterns, we also illustrated how to
``correct'' for this feature by appending the 
dark solitonic pattern on top of a nonlinear periodic wave background. 

Naturally, these types of studies pave the way for further explorations. It would be interesting to examine whether such modulated waveforms can have
any bearing in experimental settings of relevance to
the AA model. More practically, however, one can
also focus the current studies on the case of 
multiple solitary waves and
their interactions, as well as to explore more systematically
the stability of the nonlinear periodic patterns, as
has been done in the context of the KdV and related 
dispersive partial differential equations~\cite{bottman}. One can also envision variations of the present setting to variants of the AA model such as the one where the magnetic field has two components and the electric field has one component, as, e.g., in the study of~\cite{Abbas2020}. Generalizations of the present settings in higher (e.g., two) dimensions would also be of interest. 
Moreover, we note in passing that experimental observation of
Alfv{\'e}n and ion-acoustic waves (among others) is becoming 
experimentally more accessible as confirmed by data collected, e.g., by
the Radio Receiver Instrument of the Swarm-E satellite~\cite{Bernhardt2023}.
As such, associated observations of plasma waves generated, e.g., by
charged space objects expand upon the need for nonlinear models
and their characterization/description of such nonlinear wave propagation.
Such studies are currently under consideration and will be reported in future publications.


\bibliographystyle{unsrt}  
\bibliography{aa_biblio}       

\end{document}